\documentclass [10pt]{article}
\usepackage{amsfonts, epsfig}
\usepackage{multicol}
\addtocounter{page}{0}
\begin{document}

\begin{center}
\Large{\textsf{\textbf{ Yang-Mills redux}}}\\
\normalsize
\vskip .2in

\noindent
\textsf{\textup{Samuel L. Marateck}}\\
\noindent
\textsf{Courant Institute of Mathematical Sciences}\\
\noindent
\textsf{New York University}\\
\noindent
\textsf{New York, N.Y. 10012}\\
\noindent
\textsf{email:marateck@cs.nyu.edu}\\

\vskip .2in

{\bf Abstract}
\end{center}

\begin{quotation}
\noindent
It is noted that a given pairing of the phase factor and gauge
transformation to retain gauge symmetry is not unique. In their seminal paper,
when Yang and Mills (YM) discuss the phase factor - gauge transformation
relationship, they cite Pauli's review paper. It is interesting that although
Pauli in that paper presents the electromagnetic field strength in terms of a
commutator, for whatever reason YM did not extrapolate the commutator's use
to obtain the Yang-Mills field strength -- they obtained it by trial and
error. Presented is a derivation of this field strength using the commutator
approach detailing how certain terms cancel each other. Finally, the Yang-Mills
field transformation is derived in a slightly different way than is
traditionally done. 
\end{quotation}

\normalfont

\noindent
\section{Introduction}

This is an addendum to the article on differential geometry and Feynman
diagrams that appeared in the Notices of the American Mathematical Society
(Marateck 2006). It expands on some of the topics covered in the original
article.

\noindent
\section{Gauge theory}

Weyl introduced as a phase factor (Weyl 1929) an exponential in which the phase
$\alpha$ is preceded by the imaginary unit $i$, e.g., $e^{+iq\alpha({\bf x})}$,
in the wave function for the wave equations (for instance, the Dirac equation
is $(i\gamma^\mu \partial_\mu - m)\psi = 0$).  It is here that Weyl correctly
formulated gauge theory as a symmetry principle from which electromagnetism
could be derived.  It had been shown that for a quantum theory of charged
particles interacting with the electromagnetic field, invariance under a gauge
transformation of the potentials required multiplication of the wave function
by the now well-know phase factor.  Yang  cites (Yang 1986) Weyl's gauge theory
results as reported (Pauli 1941) by Pauli as a source for Yang-Mills gauge
theory; although Yang didn't find out until much later that these were Weyl's
results. Moreover, Pauli's article did not mention Weyl's geometric
interpretation. It was only much after Yang and Mills published their article
that Yang realized the connection between their work and geometry. In fact,
in his selected papers (Yang, 2005), Yang says

\begin{quotation}
\noindent
What Mills and I were doing in 1954 was generalizing Maxwell's theory. We
knew of no geometrical meaning of Maxwell's theory, and we were not looking
in that direction.
\end{quotation}

For the wave equations to be gauge invariant, i.e., have the same form after
the gauge transformation as before, the local phase transformation $\psi({\bf
x}) \rightarrow \psi(x)e^{+i\alpha({\bf x})}$ has to be accompanied by the
local gauge transformation

 \begin{equation} {\bf A_\mu} \rightarrow {\bf A_\mu} - q^{-1} {\bf
    \partial_{\mu}\alpha({\bf x})} \end{equation}

\noindent
This dictates that the $\partial_\mu$ in the wave equations be replaced by the
covariant derivative $\partial_\mu + iqA_\mu$ in order for the ${\bf
\partial_{\mu}\alpha({\bf x})}$ terms to cancel each other.  This pair of phase
factor- gauge transformation is not unique. Another pair that retains gauge
symmetry and results in the same covariant derivative has the $q$ included in
the phase factor, i.e., $\psi({\bf x}) \rightarrow \psi(x)e^{+iq\alpha({\bf
x})}$ paired with 

 \begin{equation}{\bf A_\mu} \rightarrow {\bf A_\mu} - {\bf \partial_{\mu}\alpha({\bf\ x})}
 \end{equation}

\noindent
The fact that this pairing is not unique is not surprising since the phase
factor and gauge transformation have no physical significance.\\

\noindent
\section{Yang-Mills field strength}

Pauli, in equation (22a) of Part I of his 1941 review article (Pauli 1941)
gives the electromagnetic field strength in terms of a commutator. In
present-day usage it is

\begin{equation}[D_\mu, D_\nu] = i\epsilon F_{\mu \nu}\end{equation}

\noindent
where $D_\mu$ is the covariant derivative $\partial_\mu +i\epsilon A_\mu$.
Mathematically, equation [3] corresponds to the curvature (the field strength)
reflecting the effect of parallel transport of a vector around a closed path,
i.e., its holonomic behavior. If the field strength is zero, the vector will
return to its point of origin pointing in its original direction. In their
seminal paper (Yang 1954) Yang and Mills do not mention this relation, although
they do cite Pauli's 1941 article. They use

 \begin{equation}\psi = S\psi' \end{equation}

\noindent
where $S$ is a local isotopic spin rotation represented by an SU(2) matrix, to
obtain the gauge transformation in equation [3] of their paper

 \begin{equation} B'_\mu = S^{-1}B_\mu S + iS^{-1}(\partial_\mu S)/\epsilon \end{equation}

\noindent
They\footnote{Yang had earlier started studying this problem as a graduate
 student at the University of Chicago and derived equation (5). When he
 returned to this problem as a visitor at Brookhaven, he in collaboration with
 Mills obtained (as we will explain) the field strength.} then define the field
 strength as

 \begin{equation}F_{\mu \nu} = (\partial_{\nu}B_{\mu} -\partial_{\mu}B_{\nu}) +i\epsilon
     (B_\mu B_\nu - B_\nu B_\mu) \label{eq:F} \end{equation}

\noindent
This corresponds to Cartan's second structural equation which in differential
 geometry notation is ${\bf \Omega = dA + [A, A]}$, where $A$ is a connection
 on a principal fiber bundle.\\

They introduce equation (6) (their equation [4]) by saying

\begin{quotation}
\noindent
In analogy to the procedure of obtaining gauge invariant field strengths in the
electromagnetic case, we define (4) $F_{\mu \nu} = (\partial_{\nu}B_{\mu} -
\partial_{\mu}B_{\nu}) +i\epsilon (B_\mu B_\nu - B_\nu B_\mu)$ One easily shows
from [$B'_\mu = S^{-1}B_\mu S + iS^{-1}(\partial_\mu S)/\epsilon $] that (5)
$F'_{\mu \nu} = S^{-1}F_{\mu \nu}S$ under an isotopic gauge
transformation. Other simple functions of $B$ than (4) do not lead to such a
simple transformation property.
\end{quotation}

\noindent
Yang and Mills arrived at the field strength, equation (6), by trial and error.
They added terms to the electromagnetic part until they found the commutator
part, all the while plugging the resulting field strength into their equation
[5] for verification.

Using the Yang-Mills covariant derivative $(\partial_\mu - i\epsilon B_\mu)$
let's see how the Yang-Mills field strength is obtained from the commutator

\begin{center}
$[D_\mu, D_\nu] = (\partial_\mu - i\epsilon B_\mu)(\partial_\nu -
i\epsilon B_\nu) -$ \end{center}
 \begin{equation} 
(\partial_\nu - i\epsilon B_\nu)(\partial_\mu - i\epsilon
B_\mu) \end{equation}

\noindent
operating on the wave function $\psi$.  Note that $-\partial_\mu (B_\nu \psi)
= -(\partial_\mu B_\nu) \psi - B_\nu \partial_\mu \psi$ and $\partial_\nu
(B_\mu \psi) = (\partial_\nu B_\mu) \psi + B_\mu \partial_\nu \psi$.  So we get
a needed $- B_\nu \partial_\mu$ and a $B_\mu \partial_\nu$ term to cancel
$B_\nu \partial_\mu$ and $-B_\mu \partial_\nu$ respectively. Thus expanding
(7) we get

 \begin{center}
$\partial_\mu \partial_\nu - i\epsilon \partial_\mu B_\nu - i\epsilon
B_\mu \partial_\nu - i\epsilon B_\nu \partial_\mu - \epsilon^2 B_\mu B_\nu -
\partial_\nu \partial_\mu $ \end{center}
 \begin{equation} 
+ i\epsilon \partial_\nu B_\mu + i\epsilon B_\nu
\partial_\mu + i\epsilon B_\mu \partial_\nu + \epsilon^2 B_\nu B_\mu 
 \end{equation}

\noindent
which reduces to $i\epsilon (\partial_\nu B_\mu - \partial_\mu B_\nu) -
\epsilon^2 [B_\mu, B_\nu]$ or $[D_\mu, D_\nu] = i\epsilon F_{\mu \nu}$

\noindent
\section{The field transformation}
We present a pedagogical derivation of the gauge transformation by using the
transformation

 \begin{equation}\psi' = S\psi \end{equation}

\noindent
instead of the traditional $\psi = S\psi'$, i.e., the one Yang and Mills used. 
In order to obtain the gauge transformation in equation [3] of the Yang and
Mills paper

\begin{equation}
B'_\mu = S^{-1}B_\mu S + iS^{-1}(\partial_\mu S)/\epsilon 
\end{equation}

\noindent
requires you to use\footnote{The following can be obtained by differentiating
$S^{-1}S = I$} $\partial_\mu S^{-1} = - S^{-1} (\partial_\mu S) S^{-1}$. Thus,
the approach indicated by equation (9) is marginally more
straight-forward since it doesn't require differentiating the inverse of a
matrix. 

\noindent
The covariant derivative, $D_\mu = \partial_\mu -i\epsilon B_\mu$,
transforms the same way as $\psi$ does

 \begin{equation}D'\psi' = SD\psi \end{equation}.

\noindent
The left-hand side of equation (11) becomes

\begin{equation}(\partial_\mu -i\epsilon B'_\mu)S\psi= (\partial_\mu S)\psi + 
S\partial_\mu \psi -i\epsilon B'_\mu S \psi \end{equation}

\noindent
But (12) equals $S\partial_\mu\psi -i\epsilon SB_\mu \psi.$ Cancelling
$S\partial_\mu\psi$ on both sides we get,

 \begin{equation} (\partial_\mu S) \psi -i\epsilon B'_\mu S\psi = -i\epsilon SB_\mu \psi
\end{equation} or

 \begin{equation} B'_\mu S = SB_\mu + (\partial_\mu S)/(i\epsilon)  \end{equation} thus 

 \begin{equation} B'_\mu = SB_\mu S^{-1} -i(\partial_\mu S)S^{-1}/\epsilon
\end{equation}

\noindent
We will use $S = e^{i{\bf \alpha(x) \cdot \sigma}}$. So for $\alpha$
infintessimal, $S = 1 + i\alpha \cdot \sigma$ which produces

\begin{center}
$B'_\mu = (1 + i\alpha \cdot \sigma)B_\mu (1 - i\alpha \cdot \sigma ) $
\end{center}
\begin{equation} 
- i(1/\epsilon) \partial_\mu (1 + i\alpha \cdot \sigma ) (1 - i\alpha \cdot
\sigma) \end{equation}

\noindent
Remembering that $(a \cdot \sigma)(b \cdot \sigma) = a \cdot b + i\sigma \cdot
(a \times b)$, setting $B_\mu = \sigma \cdot b_\mu$, and since $\alpha$ is
infintessimal, dropping terms of order $\alpha^2$, we get

\begin{center}
$b'_\mu \cdot \sigma = b_\mu \cdot \sigma$\end{center}\begin{equation} 
+ i[(\alpha \cdot \sigma)(b_\mu
  \cdot \sigma), (b_\mu \cdot \sigma)(\alpha \cdot \sigma)] + (1/\epsilon)
\partial_\mu (\alpha \cdot \sigma ) \end{equation}and finally
\begin{equation} 
b'_\mu = b_\mu + 2(b_\mu \times \alpha)
+ (1/\epsilon) \partial_\mu
\alpha \end{equation}

\noindent
which (because our S is the inverse of Yang-Mills' S) is equation [10]
in the Yang-Mills paper. \\

\noindent
\section*{Acknowledgements}
\noindent
The author thanks Ed Osinski and Alana Libonati for their help in typesetting
these papers.

\noindent
\section*{References}

\noindent
Marateck, Samuel L., 2006. Notic. Amer. Math. Soc. {\bf 53}
744.\\
Pauli, W., 1941. Rev. Mod. Physics. {\bf 13} 203.\\
\noindent
Weyl, Hermann, 1929. Zeit. f. Physic. {\bf 330} 56.\\
\noindent
Yang, C. N. and Mills, R. L., 1954. Phys. Rev. {\bf 96} 191.\\
\noindent
Yang, C.N., 1986 in {\it Hermann Weyl's contribution to
Physics}, in {\it Hermann Weyl:1885- 1985}, ed. Chandrasekharan, K. (Springer-Verlag).\\
\noindent
Yang, C.N., 2005 in {\it Selected Papers (1945-1980) With Commentary}, World
Scientific. p74.

\end{document}